\documentclass[reprint,showpacs,preprintnumbers,amsmath,amssymb,aps,pre]{revtex4-1}

\usepackage{graphicx}
\usepackage{dcolumn}
\usepackage{bm}
\usepackage[OT4]{fontenc} 
\newcommand{\rozmiar}{0.325}

\begin{document}

\title{Stable and metastable phases  in the atomic limit of \\the extended Hubbard model
with intersite density--density interactions}
\author{Konrad Kapcia}%
    \email{corresponding author; e-mail: \url{kakonrad@amu.edu.pl}}
\affiliation{Electron States of Solids Division, Faculty of Physics, Adam Mickiewicz University, Umultowska 85, 61-614 Pozna\'n, Poland
}%
\author{Stanis\l{}aw Robaszkiewicz}%
\affiliation{Electron States of Solids Division, Faculty of Physics, Adam Mickiewicz University, Umultowska 85, 61-614 Pozna\'n, Poland
}%

\date{May 1, 2011}

\begin{abstract}
We have studied a~simple effective model of charge ordered insulators. The tight binding Hamiltonian consists of the effective on-site interaction $U$ and the intersite density-density interaction $W_{ij}$ (both: nearest-neighbor and next-nearest-neighbor).
In the analysis of the phase diagrams and thermodynamic properties of this model we have adopted the variational approach, which treats the on-site interaction term exactly and the intersite interactions within the mean-field approximation.
Our investigations of the general case (as a function of the electron concentration $n$) have shown that
the system exhibits various critical behaviors  including among others bicritical, tricritical, critical-end and isolated critical points.
In this report we concentrate on the metastable phases and transitions between them. One finds that the first- and second order transitions between metastable phases can exist in the system. These transitions occur in the neighborhood of first as well as second order transitions between stable phases. For the case of on-site attraction the regions of metastable homogeneous phases occurrence inside the ranges of phase separated states stability have been also determined.
\end{abstract}

\maketitle


\section{Introduction}

There is intense research in the field of electron charge orderings phenomena due to their relevance for a broad range of important materials such as manganites, cuprates, magnetite, several nickel, vanadium and cobalt oxides, heavy fermion systems and numerous organic compounds (Refs.~\cite{MRR1990,IFT1998,DHM2001,SHF2004,F2006} and references therein).

The effective Hamiltonian of an~electron system on the lattice in the zero bandwidth limit considered in this report can be written in the following form:
\begin{eqnarray}
\label{row:1} \hat{H}  & = & U\sum_i{\hat{n}_{i\uparrow}\hat{n}_{i\downarrow}} + \frac{W_{1}}{2}\sum_{\langle i,j\rangle_1}{\hat{n}_{i}\hat{n}_{j}} +\\
& +& \frac{W_{2}}{2}\sum_{\langle i,j\rangle_2}{\hat{n}_{i}\hat{n}_{j}} - \mu\sum_{i}{\hat{n}_{i}}, \nonumber
\end{eqnarray}
where $\hat{c}^{+}_{i\sigma}$ denotes the creation operator of an electron with spin $\sigma$ at the site~$i$, $\hat{n}_{i}=\sum_{\sigma}{\hat{n}_{i\sigma}}$, $\hat{n}_{i\sigma}=\hat{c}^{+}_{i\sigma}\hat{c}_{i\sigma}$,
$U$~is the on-site density interaction,
$W_{1}$ and $W_{2}$ are the intersite density-density interactions between nearest neighbors (nn)
and next-nearest neighbors (nnn), respectively. These interactions will be treated as the effective ones and will be assumed to include all the possible contributions and renormalizations. $\mu$ is the chemical potential, depending on the concentration of electrons
%
\mbox{$n = \frac{1}{N}\sum_{i}{\left\langle \hat{n}_{i} \right\rangle}$},
%
with \mbox{$0\leq n \leq2$} and $N$ is the total number of lattice sites. Our denotations: \mbox{$n_Q=\frac{1}{2}(n_A-n_B)$}, \mbox{$n_\alpha=\frac{2}{N}\sum_{i \in \alpha}\langle \hat{n}_i \rangle$},
and \mbox{$\alpha=A,B$} labels the sublattices. \mbox{$W_0 = z_1W_1+z_2W_2$}, \mbox{$W_Q = -z_1W_1+z_2W_2$}, where $z_1$ and $z_2$ are the number of nn and nnn, respectively.

We have performed extensive study of the phase diagrams of the model (\ref{row:1}) for \mbox{$W_1>0$} and arbitrary $n$ \mbox{\cite{R1979,MRC1984,KR2011,KKR2010,KK2009}}. Depending on the values of model parameters the system can exhibit not only several homogeneous charge ordered (CO) phases and nonordered (NO) phase, but also various phase separated (PS) states (PS1: CO-NO, PS2: CO-CO, PS3: NO-NO) \mbox{\cite{KR2011,KKR2010,KK2009,BT1993}}, in which two domains with different concentration exist (coexistence of two homogeneous phases). However, the behaviors of metastable phases occurring in  model (\ref{row:1}) have not been analyzed till now.

In the analysis we have adopted a~variational approach (VA), which treats the on-site interaction term ($U$) exactly and the intersite interactions ($W_{ij}$) within the mean-field approximation (MFA). One obtains two equations for $n$ and $n_Q$, which are solved self-consistently. Explicit forms of equations for the free energy and other thermodynamical properties are derived in Ref.~\cite{KR2011}.
$n_Q$ is non-zero in the charge-ordered phase, whereas in  the nonordered phase \mbox{$n_Q=0$}. Only the two-sublattice orderings on the alternate lattices are considered in this report.

In present report we will concentrate on the possibility of metastable phases occurrence on the phase diagrams of model considered.

\begin{figure*}
    \centering
    \includegraphics[width=\rozmiar\textwidth]{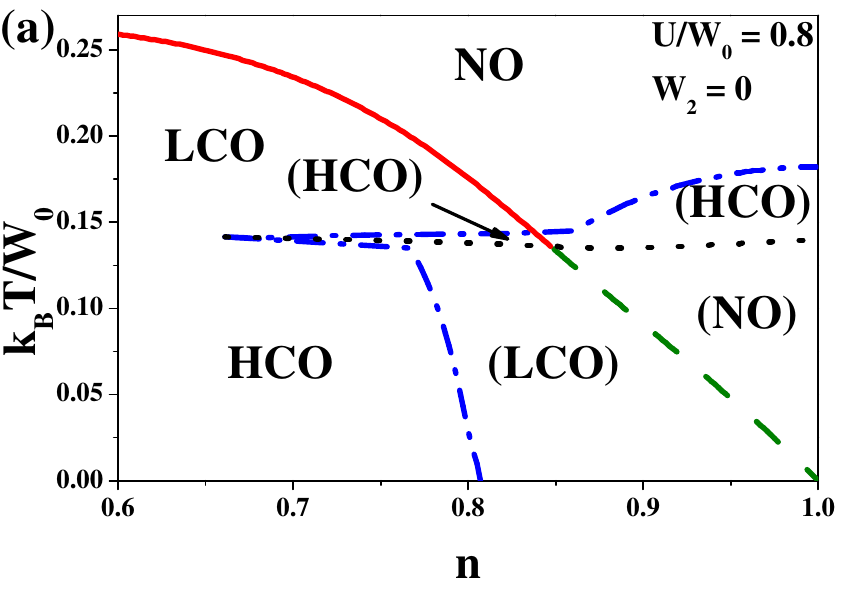}
    \includegraphics[width=\rozmiar\textwidth]{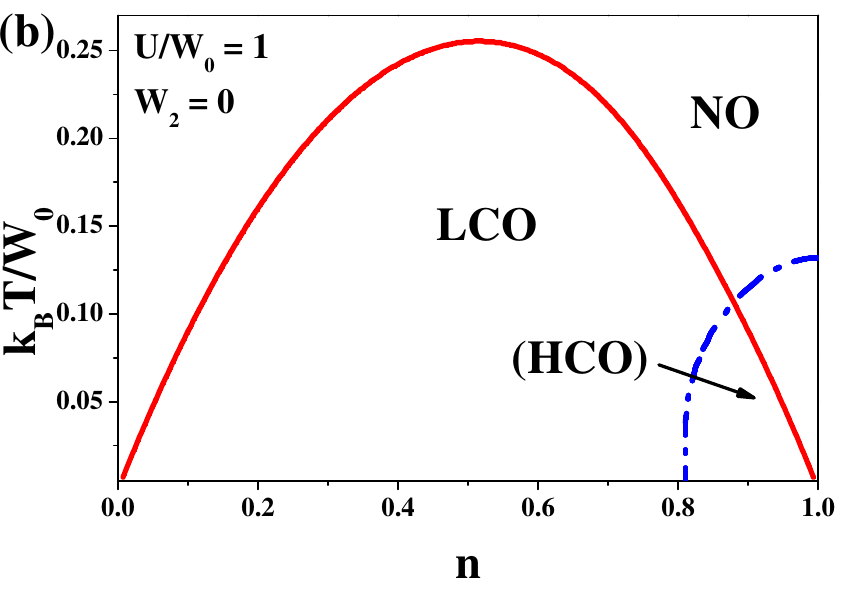}
    \includegraphics[width=\rozmiar\textwidth]{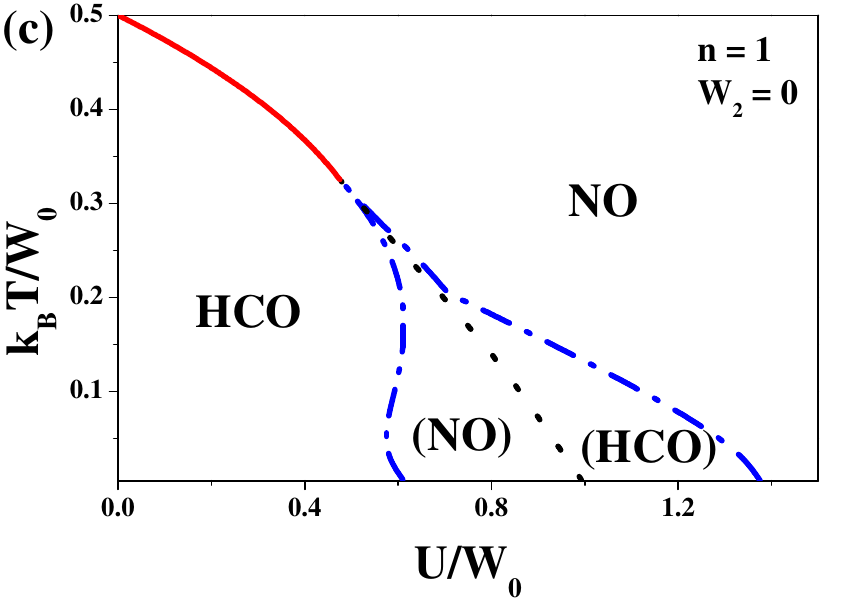}
    \caption{Phase diagrams for \mbox{$W_1>0$}, $W_2=0$ and: (a)~\mbox{$U/W_0=0.8$}, (b)~\mbox{$U/W_0=1.0$}, (c)~\mbox{$n=1$}. Dotted and solid lines denote first and second order transitions between stable phases. Dashed-dotted lines denote the boundaries of metastable phase occurrence (names of metastable phases in brackets). Dashed line (panel (a)) denotes second order boundary between metastable phases.}
    \label{rys:fig1}
\end{figure*}
\begin{figure*}
    \centering
    \includegraphics[width=\rozmiar\textwidth]{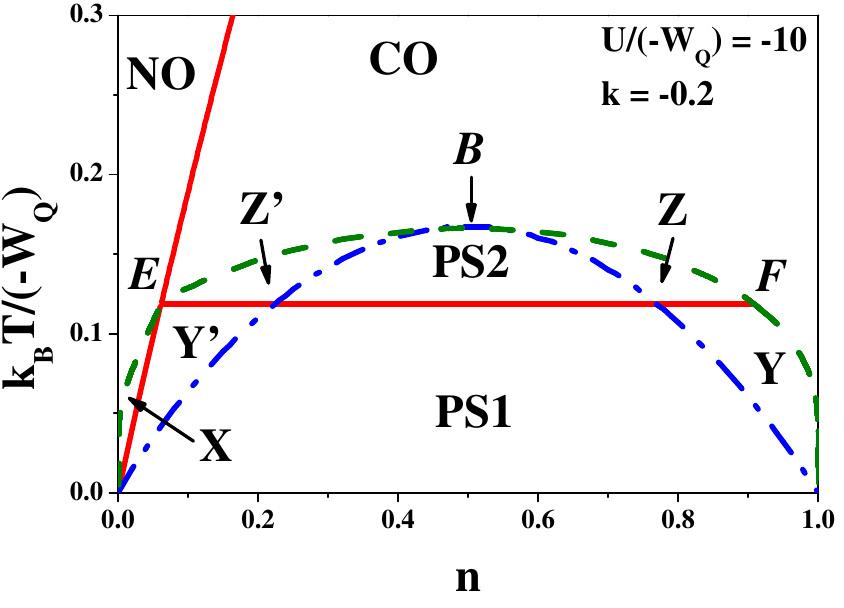}
    \includegraphics[width=\rozmiar\textwidth]{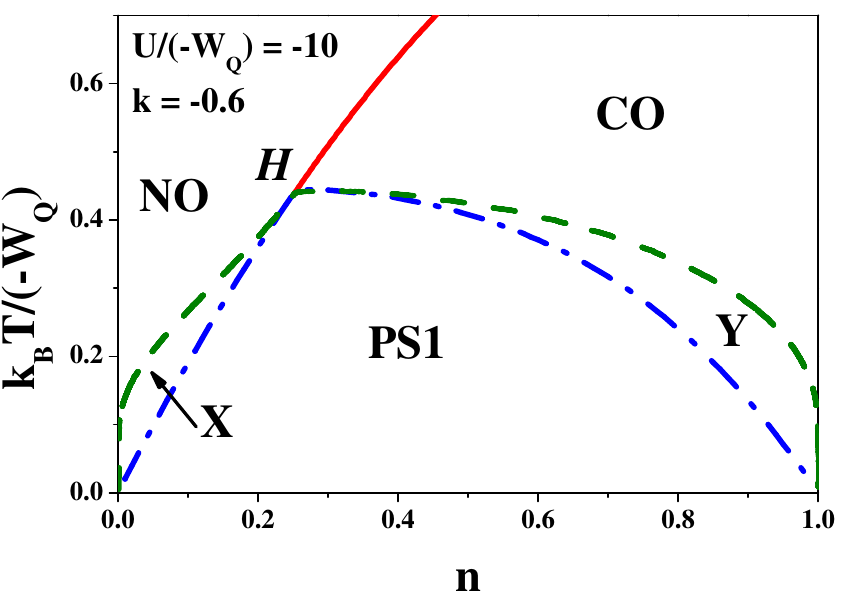}
    \includegraphics[width=\rozmiar\textwidth]{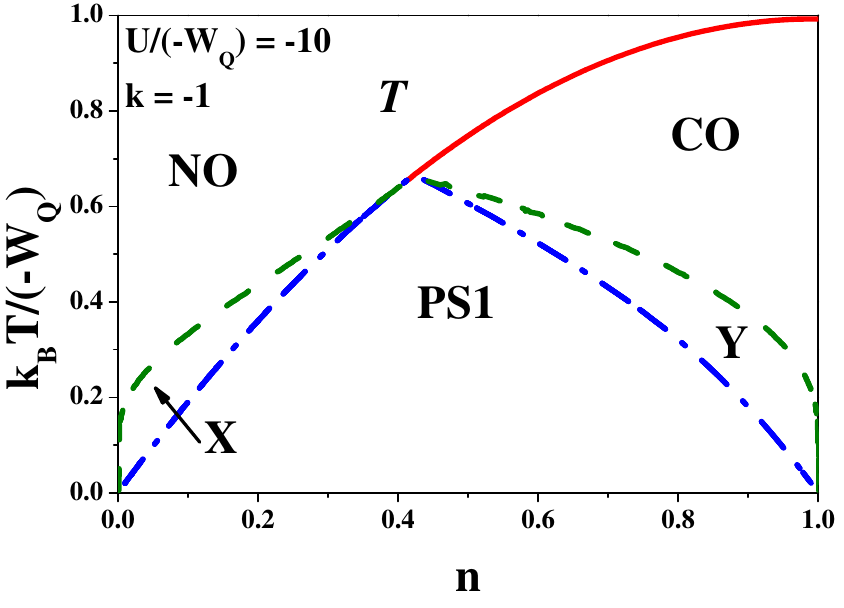}
    \caption{Phase diagrams for \mbox{$U/(-W_Q)=-10$}, \mbox{$W_1>0$} and \mbox{$k=z_2W_2/z_1W_1=-0.2,-0.6,-1.0$} (as labeled). CO denotes now the HCO phase. Solid and dashed lines indicate second order and ``third order'' transitions, respectively. Below dashed-dotted lines all homogeneous phases are unstable. Details in text.}
    \label{rys:fig2}
\end{figure*}

\section{Results and discussion}

\subsection{$W_1>0$, $W_2=0$}

For \mbox{$W_1>0$} and \mbox{$W_2=0$} the system exhibits  a tricritical line, a critical end point line and a line of isolated critical points \cite{MRC1984}. The CO-NO transition can be second order as well as first order. Two different CO phases (i.e. LCO and HCO) are separated by first order line.

In Fig.~\ref{rys:fig1} we presents a few particular phase diagrams involving metastable phases. It is quite obvious that metastable phases are present in the neighborhood of first order (\mbox{HCO--LCO} and \mbox{HCO--NO}) transitions (such region is very narrow for the \mbox{HCO--LCO} transition).  Above the first order transition temperature the phase, which was stable below the transition temperature, is metastable, and inversely, below the transition temperature  the phase, which was stable above the transition temperature, is metastable.  However, one should notice that  second order \mbox{LCO--NO} transition occurs between two metastable phases with increasing temperature connected with continuous change of charge-order parameter in metastable phases (Fig.~\ref{rys:fig1}a, \mbox{$U/W_0=0.8$}). Such transition between metastable phases occurs in the higher energy branch of solutions, whereas the lowest energy solution is the HCO phase. Other interesting feature of the model is that in the vicinity  of second order \mbox{LCO--NO} transition for \mbox{$n>0.8$} the HCO phase is metastable (Fig.~\ref{rys:fig1}b, \mbox{$U/W_0=1$}). This behavior is connected with \mbox{HCO--NO} transition occurring for \mbox{$U/W_0<1$} (cf. Fig.~\ref{rys:fig1}c).

Let us stress that we found all MFA solutions of the model considered. Thus metastable phases occur only in the regions explicitly denoted on the phase diagrams. In other regions there are no metastable phases - only one (stable) solution exists.

\subsection{$W_1>0$, $W_2<0$}

In such range of model parameters the system can exhibits not only several CO phases, but also various phase separated states: PS1 and PS2 \cite{KR2011,KKR2010}. Examples of the $k_BT$~vs.~$n$ phase diagrams evaluated for strong on-site attraction \mbox{$U/(-W_Q)=-10$}, \mbox{$W_1>0$} and various ratios of \mbox{$k=z_2W_2/z_1W_1< 0$} are shown in Fig.~\ref{rys:fig2}.
A~transition between homogeneous phase and PS state is symbolically named as a~``third order'' transition. During this transition a~size of one domain in the PS state decreases continuously to zero at the~transition temperature.
The CO and NO phases are separated by the second order transition line and for \mbox{$k=0$} no metastable phases occur. If \mbox{$k<0$} in the ranges of PS stability the homogeneous phases can be metastable (if \mbox{$\partial \mu/\partial n >0$}) as well as unstable (if \mbox{$\partial \mu/\partial n <0$}).

For \mbox{$k<-0.6$} the PS1 state occurs on the phase diagram
and the critical point for the phase separation (denoted as $T$) lies on the second order line \mbox{CO--NO}. As \mbox{$k \rightarrow -\infty$} the \mbox{$T$-point} occurs at \mbox{$n=1$} and the homogeneous CO phase does not exist beyond half-filling. If \mbox{$k=-0.6$} $H$-point is present on the phase diagram and the system changes a~tricritical behavior (for \mbox{$k<-0.6$}) into a~bicritical behavior (for \mbox{$0>k>-0.6$}). In the ranges of PS1 stability the NO phase (in region $X$) and the CO phase (in region $Y$) are metastable. Below dashed-dotted lines all homogeneous phases considered (CO as well as NO) are unstable (i.e. \mbox{$\partial \mu/\partial n <0$} in all homogeneous solutions).

When \mbox{$-0.6<k<0$} a transition between PS state and homogeneous phase takes place at low temperatures, leading first to phase separation into two coexisting CO phases (PS2), while at still lower temperatures CO and NO phases coexist (PS1). The critical point (denoted as $B$) for this phase separation is located inside the CO phase. The \mbox{$E$-$F$} solid line is associated with continuous transition between two different PS states (\mbox{PS1--PS2}, the second order \mbox{CO--NO}  transition occurs in the domain with lower concentration).
Similarly as for \mbox{$k\leq-0.6$},  for \mbox{$-0.6<k<0$} the NO phase (in region $X$) or the CO phase (in regions $Y$ and $Y'$) are metastable in the ranges of PS1 stability. One should notice that second order transition \mbox{CO--NO} between metastable phases occurs (the solid line between regions $X$ and $Y'$ in Fig.~\ref{rys:fig2} for \mbox{$k=-0.2$}).  At higher temperatures, in the ranges of PS2 stability only the CO phase can be metastable (in regions $Z$ and $Z'$). Below dashed-dotted line all homogeneous phases considered are unstable.

For larger values of \mbox{$U/(-W_Q)$} (especially if \mbox{$U/(W_Q)>0$} it could be possible that more than one metastable phase exist in ranges of PS states occurrence, however we do not analyze it in this report.

\subsection{$W_1<0$, $W_2=0$}

\begin{figure}
    \centering
    \includegraphics[width=\rozmiar\textwidth]{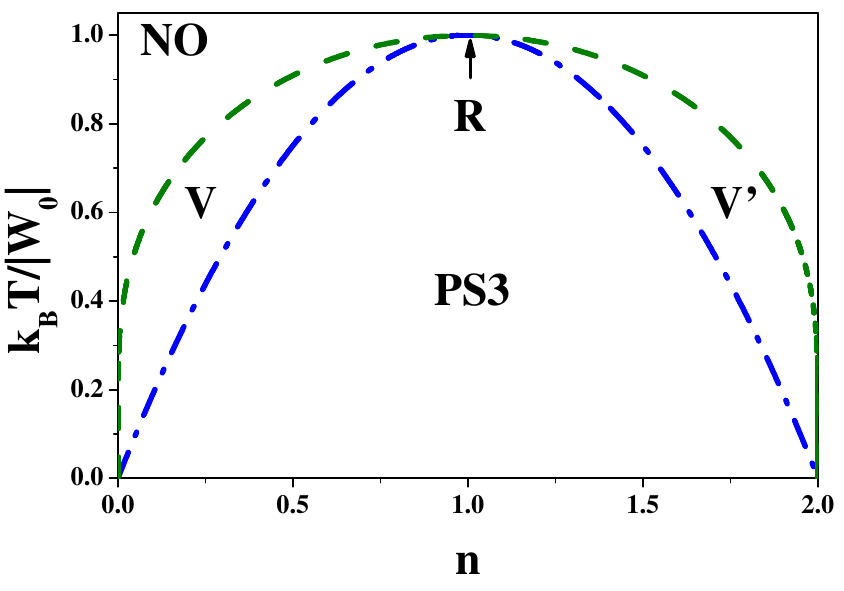}
    \vspace{-0.3cm}
    \caption{Phase diagram for \mbox{$U/(-W_Q)=-10$}, \mbox{$W_1<0$} and \mbox{$W_2=0$}. Dashed line indicates the \mbox{PS3--NO} transitions.  Below dashed-dotted line the homogeneous NO phase is unstable.}
    \label{rys:fig3}
\end{figure}

For the case \mbox{$W_1<0$} (\mbox{$W_2=0$}) the model (\ref{row:1}) (\mbox{$W_2=0$}) exhibits a phase separation \mbox{NO-NO} (electron droplets state -- PS3) at low temperatures~\cite{BT1993}. In this PS state different spatial non-ordered regions have different average electron concentrations. In such a~case, at higher temperatures only the homogeneous NO phase occurs. The phase diagram for \mbox{$U/|W_0|=-10$} and \mbox{$W_1<0$} involving metastable phases is shown in  Fig.~\ref{rys:fig3}. One can notice that the homogeneous NO phase is metastable in regions $V$ and $V'$. The line restricting (meta-)stability of the NO is tangent to the PS3-NO boundary in the $R$-point ($R$ is a~bicritical point).
Below dashed-dotted line the homogeneous  NO phase is unstable (i.e. \mbox{$\partial \mu/\partial n <0$} in the NO phase).

\section{Conclusions}

In this report, we have presented  some particular phase diagrams of the extended Hubbard model with intersite density-density interactions in the zero-bandwidth limit.
We have found that the first- and second order transitions between metastable phases can exist in the system. These transitions occur in the neighborhood of first as well as second order transition between stable phases. We have also determined the regions of metastable homogeneous phases occurrence inside the ranges of phase separated states stability for the case of on-site attraction.


\begin{acknowledgments}
K.~K. would like to thank the European Commission and Ministry of Science and Higher Education (Poland) for the partial financial support from European Social Fund -- Operational Programme ``Human Capital'' -- POKL.04.01.01-00-133/09-00 -- ``\textit{Proinnowacyjne kszta\l{}cenie, kompetentna kadra, absolwenci przysz\l{}o\'sci}''.
\end{acknowledgments}

\end{document}